\documentclass[aps,prl,twocolumn,superscriptaddress]{revtex4-1}

\usepackage{graphicx}
\usepackage{dcolumn}
\usepackage{bm}
\usepackage{epstopdf}
\usepackage{float}
\usepackage{amsmath}
\usepackage{color}
\usepackage{comment}
\graphicspath{{./images/}}

\begin{document}


\title{Mediated interactions between ions in quantum degenerate gases}
\author{Shanshan Ding}
\affiliation{Center for Complex Quantum Systems, Department of Physics and Astronomy, Aarhus University, Ny Munkegade, DK-8000 Aarhus C, Denmark. }
\author{Michael Drewsen}
\affiliation{Center for Complex Quantum Systems, Department of Physics and Astronomy, Aarhus University, Ny Munkegade, DK-8000 Aarhus C, Denmark. }
\author{Jan J.\  Arlt}
\affiliation{Center for Complex Quantum Systems, Department of Physics and Astronomy, Aarhus University, Ny Munkegade, DK-8000 Aarhus C, Denmark. }
\author{G. M. Bruun}
\email[]{bruungmb@phys.au.dk}
\affiliation{Center for Complex Quantum Systems, Department of Physics and Astronomy, Aarhus University, Ny Munkegade, DK-8000 Aarhus C, Denmark. }
\affiliation{Shenzhen Institute for Quantum Science and Engineering and Department of Physics, Southern University of Science and Technology, Shenzhen 518055, China}




\date{\today}

\begin{abstract}
We explore the interaction between two trapped ions mediated by a surrounding quantum degenerate Bose or Fermi gas. Using perturbation theory valid for
weak atom-ion interaction, we show  analytically that the  interaction mediated by a Bose gas  has a  power-law behaviour for large distances whereas
it has a Yukawa form for intermediate distances. For a Fermi gas, the mediated interaction is given by a power-law for large density and by a
 Ruderman-Kittel-Kasuya-Yosida form for low density. For strong atom-ion interactions, we use a diagrammatic theory to demonstrate that the mediated interaction can be a significant
 addition to  the bare Coulomb  interaction between the ions,  when an atom-ion 
 bound state is close to threshold.
 Finally, we show that the induced interaction leads to substantial and observable shifts in the ion phonon frequencies.
 \end{abstract}
\maketitle

Mediated interactions play a crucial role for our understanding of nature.
 They  are central to Landau's widely used quasiparticle theory~\cite{Baym1991}, and all interactions are mediated by gauge bosons
at a fundamental level~\cite{weinberg_1995}. Conventional and high temperature superconductivity is caused by interactions mediated by lattice phonons or
spins~\cite{SCALAPINO1995329}, and  effective interactions between  bosons mediated by fermions  in an
atomic gas have been observed~\cite{DeSalvo:2019uc,Edri2020,Fritsche2021}.
The exquisite single particle control  of trapped ions combined with the great flexibility of  atomic gases makes  hybrid atom-ion systems a
promising new platform to systematically explore mediated interactions~\cite{tomza2019cold}. In such a system, the induced interaction between ions mediated by the surrounding atoms
 can be  controlled  using the many experimentally tunable parameters of atomic gases, and accurately detected with high precision ion spectroscopy.
 Remarkable experimental progress has been reported  regarding  atom-ion
collisions, sympathetic cooling,  molecular physics~\cite{grier2009observation,zipkes2010trapped,harter2012single,ratschbacher2012controlling,sikorsky2018spin,feldker2020buffer,schmidt2020optical},
Rydberg atom-ion mixtures~\cite{Secker2016,Secker2017}, and
 mobile ions in a Bose-Einstein condensate (BEC)~\cite{kleinbach2018ionic,dieterle2020inelastic,dieterle2021transport,Veit2021}.
Recently, atom-ion Feshbach resonances were observed, opening up the
 possibility to tune the interaction strength~\cite{weckesser2021observation}. Theoretically, the
 properties of a static ion in a BEC were explored~\cite{massignan2005static}, and mobile ions have been predicted to form
quasiparticles when they are immersed in a BEC~\cite{casteels2011polaronic,astrakharchik2021ionic, christensen2021charged}
 or a  Fermi gas~\cite{christensen2021mobile}, which are charged analogues of neutral Bose and Fermi polarons~\cite{Jorgensen2016,Hu2016,Ardila2019,Yan2020,Schirotzek2009,Kohstall2012,Koschorreck12,Massignan2014,Scazza2017,Fritsche2021}.

 Here, we  investigate the  interaction between two  ions mediated by  a  surrounding BEC or  Fermi gas. Using a combination of analytical
and numerical calculations, we show that  such hybrid atom-ion systems can be used to accurately  probe mediated interactions  with
ion phonon spectroscopy, and that they can be tuned to new regimes using the flexibility of atomic gases.

\begin{figure}[h]
\centering
\includegraphics[width=0.45\textwidth]{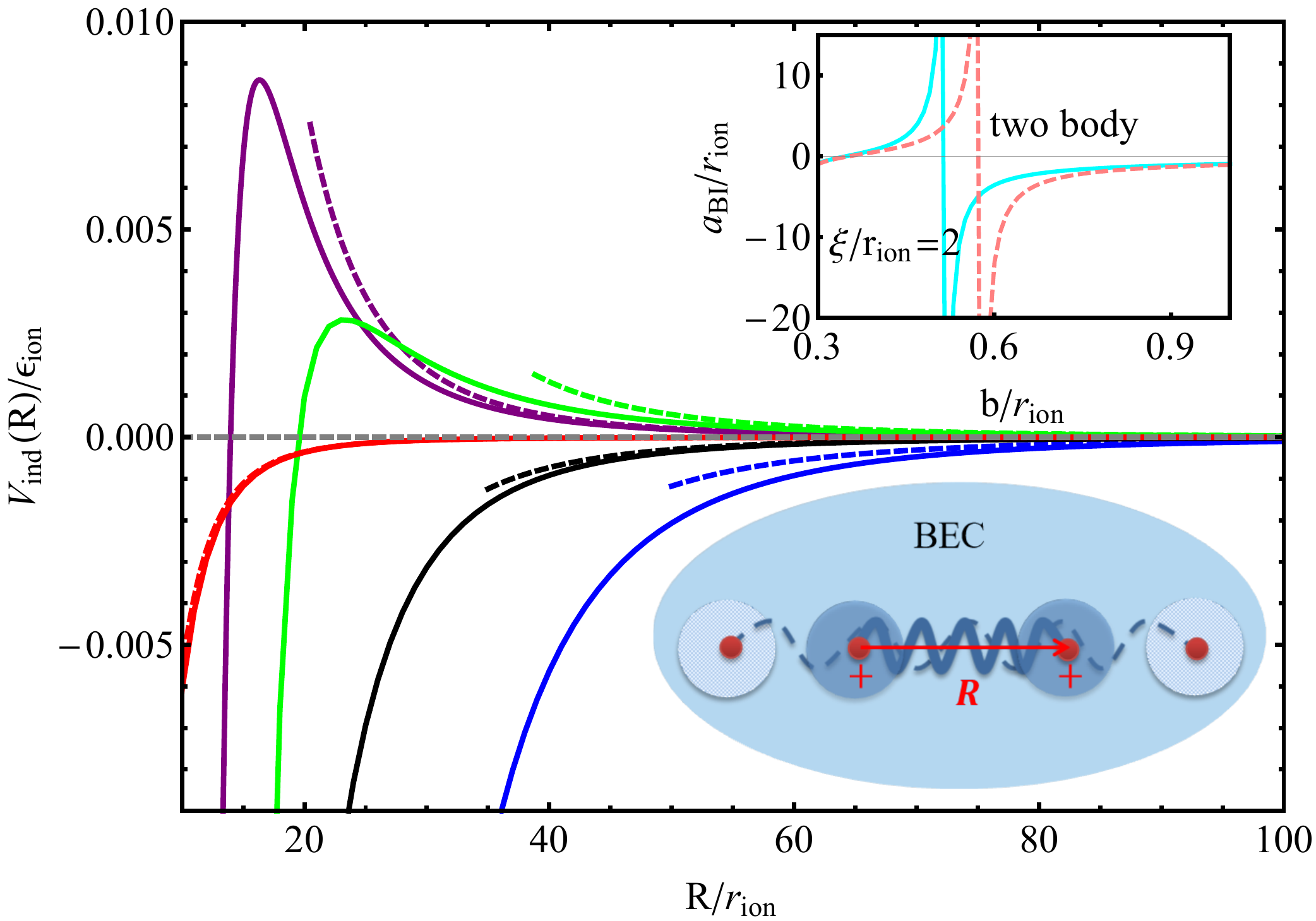}
\caption{Cartoon: Two ions interact via density modulations in the surrounding BEC. This changes  the out-of-phase phonon frequency in a linear rf trap. Main plot:  The induced interaction $V_{\mathrm{ind}}(R)$ in units of $\epsilon_\text{ion}=1/2mr_\text{ion}^2$
between two ions mediated by a BEC with density
$nr_{\mathrm{ion}}^3=1$, healing length $\xi/r_\text{ion}=2$, and  $b/r_\text{ion}=1.4$ (red),
$0.53$ (black), $0.52$ (blue),  $0.51$ (green), and $0.5$ (purple) corresponding to the atom-ion in-medium scattering lengths
$r_{\mathrm{ion}}/a_{BI}=-1.641, -0.054, -0.014, 0.029, 0.076$, respectively.
The solid lines are the numerical results obtained from Eqs.~\eqref{VindStrong}, while the dashed lines
are the analytical results given by Eqs.~\eqref{LongrangeBECWeak} and \eqref{VBoseStrongLongrange}.
Inset: The atom-ion scattering length $a_{BI}$ in the BEC  medium (solid line) and a vacuum $a_{BI,v}$ (dashed line).
}
\label{Fig-setup}
\end{figure}

\paragraph{Model.--}
We consider two static ions separated by   $\mathbf{R}$ in a spatially homogenous three-dimensional quantum degenerate gas consisting of identical bosonic or fermionic atoms of mass $m$ and density $n$ at zero temperature, see Fig.~\ref{Fig-setup}. 
The Hamiltonian  is
\begin{equation}\begin{aligned}
\hat H=&\sum_\mathbf{k}\frac{k^2}{2m}\hat c_\mathbf{k}^\dagger\hat c_\mathbf{k}+\frac{g}{2}\sum_{\mathbf{k},\mathbf{k}',\mathbf{q}}\hat c^\dagger_{\mathbf{k}+\mathbf{q}}
\hat c^\dagger_{\mathbf{k}'-\mathbf{q}}\hat c_{\mathbf{k}'}\hat c_\mathbf{k}\\
&+\sum_{\mathbf{k},\mathbf{q}}V_q\hat c^\dagger_{\mathbf{k}+\mathbf{q}}\hat c_{\mathbf{k}}\left(1+e^{-i\mathbf{q}\cdot\mathbf{R}}\right),
\end{aligned}
\label{Eq-H-two-ions}\end{equation}
where $\hat c^\dagger_\mathbf{k}$ creates an atom with  momentum $\mathbf{k}$ and
$\hat c_\mathbf{k}\hat c^\dagger_{\mathbf{k}'}\mp\hat c^\dagger_{\mathbf{k}'}\hat c_\mathbf{k}=\delta_{{\mathbf k},{\mathbf k}'}$ for bosons/fermions as usual.
The Fourier transform  of the atom-ion interaction is $V_q$  and
$\hbar$ and the system volume are set to  unity.  We define $g=4 \pi a_{BB}/m$ with $a_{BB}$ the atom-atom scattering length and
  $n^{1/3}a_{BB}\ll1$,  so  that interaction effects can be described using Bogoliubov theory for bosons. The atom-atom interaction plays no
    role for a single-component Fermi gas due to  Pauli exclusion.

The electric field from the ions gives rise to a long range atom-ion interaction $-\alpha_\text{atom}/2r^4$
with $\alpha_\text{atom}$ the atom polarizability~\cite{tomza2019cold}.
 The corresponding characteristic length scale $r_{\mathrm{ion}}=\sqrt{2m\alpha}$ where
$\alpha=\alpha_\text{atom}/2$ can easily be of  the same order of magnitude as the average inter-atom distance, so
 it is crucial to include the long range  $1/r^4$ tail explicitly in our theory~\cite{massignan2005static}. To include the short range repulsion due to the overlap between the
atom and ion electron clouds, we use the effective interaction ~\cite{krych2015description}
\begin{equation}
V\left(r\right)=-\frac{\alpha}{\left(r^2+b^2\right)^2}\frac{r^2-c^2}{r^2+c^2},
\label{Potential}
\end{equation}
 where $b\sim{\mathcal O}(r_\text{ion})$ gives the inverse depth of the potential and $c\sim a_0\ll r_\text{ion}$
 the transition point  between $V(r)>0$ and $V(r)<0$. While the full interaction
 potential is complicated and supports many bound states~\cite{tomza2019cold}, their energy separation
 is much larger than the relevant energies for the present many-body problem. It is therefore sufficient to
 use Eq.~\eqref{Potential} with  $b$ and $c$ determined so that it
recovers the atom-ion scattering length and the energy of the highest bound state of a given atom-ion combination.
Here, we fix  $c=0.0023r_{\mathrm{ion}}$ in our numerical calculations and
vary $b$, thereby tuning the strength and presence of 
maximally one  bound state of Eq.~\eqref{Potential}. This mimics the tuning of the energy of the highest bound state
in a real atom-ion system, which gives rise to the recently observed Feshbach resonances~\cite{weckesser2021observation}.

\begin{figure}[h]
\includegraphics[width=0.47\textwidth]{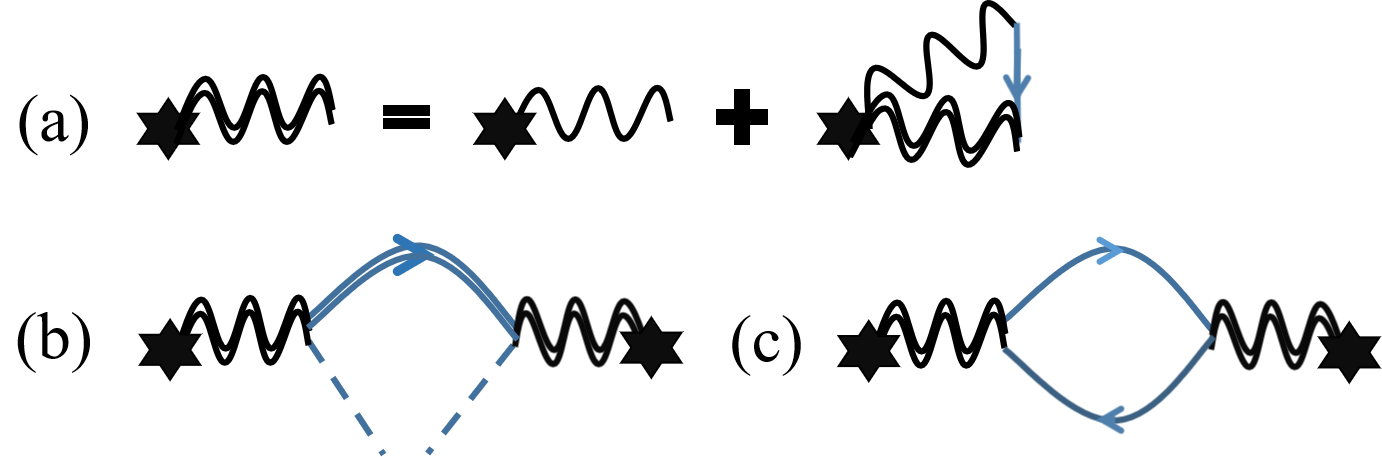}
\caption{(a) Diagrams for the atom-ion scattering matrix. A  $\ast$ is an
 ion, a wavy line is the atom-ion interaction, and a blue line is the
atom Green's function. (b) The induced interaction mediated by a sound mode (double line) in a BEC. Dashed lines are condensate atoms. (c)  The induced interaction mediated by a particle-hole excitations in a Fermi gas. }
\label{FeynFig}
\end{figure}

\paragraph{Atom-ion scattering.--} Consider first the
 scattering of an atom on a static ion. The scattering matrix  obeys
\begin{equation}
{\mathcal T}(\mathbf{p}', \mathbf{p}; \omega)=
V_{\mathbf{p}'-\mathbf{p}}+\sum_\mathbf{k}{\mathcal T}(\mathbf{p}', \mathbf{k}; \omega)G(\mathbf{k}, \omega)V_{\mathbf{k}-\mathbf{p}}
\label{ScatteringMatrix}
\end{equation}
in the ladder approximation, see  Fig.~\ref{FeynFig}(a). Here,  $\mathbf p$ ($\mathbf p'$) is the momentum of the in-coming (out-going) atom with energy $\omega$ and
$G(\mathbf{k}, \omega)$ is the atom Green's function \cite{bruus2004many}. Equation \eqref{ScatteringMatrix} is exact  in the case of vacuum scattering
where $G(\mathbf{k}, \omega)=1/(\omega-{\mathbf k}^2/2m)$ and  the atom-ion scattering length can be obtained as
$a_{BI,v}=m\mathcal{T}_\text{vac}(0,0;0)/2\pi$.
When the scattering occurs in the BEC, we use the Bogoliubov Green's function
$G(\mathbf{k},\omega)=u_\mathbf{k}^2/(\omega-E_\mathbf{k})-v_\mathbf{k}^2/(\omega+E_\mathbf{k})$
in Eq.~\eqref{ScatteringMatrix}, where $E_\mathbf{k}=\sqrt{\epsilon_\mathbf{k}^2+2ng\epsilon_\mathbf{k}}$  is the  excitation spectrum and
$v_\mathbf{k}^2=u_\mathbf{k}^2-1=\left[\left(\epsilon_\mathbf{k}+ng\right)/E_\mathbf{k}-1\right]/2$. Then we define the scattering length in the BEC as
 $a_{BI}=m\mathcal{T}(0,0;0)/2\pi$. In the inset of Fig.~\ref{Fig-setup}, we plot the scattering length $a_{BI}$ as a function of $b$
  in a BEC with     
  healing length $\xi=1/\sqrt{8\pi n a_{BB}}=2r_\text{ion}$. We also plot the scattering length $a_{BI,v}$ in a vacuum  for comparison.
 The scattering diverges at $b\simeq0.5168r_{\mathrm{ion}}$ where  a bound atom-ion state emerges. We see that this is
  a smaller value of $b$ than for the vacuum case corresponding to a deeper atom-ion interaction potential, which shows that
the BEC suppresses the formation of  bound states.  When the
 scattering occurs in a Fermi gas, we use  the Fermi Green's function $G(\mathbf{k},\omega)=1/(\omega-k^2/2m+\epsilon_F)$ with $\epsilon_F=k_F^2/2m=(6\pi^2n)^{2/3}/2m$ the Fermi energy. Since scattering at the Fermi surface is the most important, we define the scattering length in a Fermi gas as
 $a_{FI}=m\text{Re} [\mathcal{T}_s(k_F,k_F;0)]/2\pi$, where the subscript $s$ represents the $s$-wave component.
The scattering matrix has poles at the bound state energies of an atom in the potential of the  ion, and   ${\mathcal T}(0,0; 0)$ diverges every time a new  bound state
appears. 
We have ${\mathcal T}_{\mathbf R}(\mathbf{p}', \mathbf{p}; \omega)={\mathcal T}(\mathbf{p}', \mathbf{p}; \omega)\exp[-i(\mathbf{p}'- \mathbf{p})\cdot {\mathbf R}]$
for  the scattering matrix of the ion at position ${\mathbf R}$.

\paragraph{BEC and weak interaction.--}
We now analyse the induced interaction between  two ions in a BEC, considering first  the case of a weak atom-ion interaction $b \gtrsim r_\text{ion}$
so that  there is no shallow two-body bound state. 
The induced interaction can then be extracted rigorously from a perturbative calculation of
 the energy shift due to the presence of the two ions. To first order, we obtain the  Hartree shift $ E_1=2nV_{q=0}$, which
 does not depend on their separation $\mathbf R$. The second order energy shift is~\cite{SM}
\begin{align}
E_2=\sum_{\mathbf q}V_q^2[1+\cos({\mathbf q}\cdot {\mathbf R})]\chi(q,0)=\tilde E_2+V_\text{ind}(R)
\label{EnergyShift}
\end{align}
with $\chi(q,0)$ the static density-density response function of the gas. The constant term $\tilde E_2$ is
 the second order energy shift coming from each ion separately,
 whereas we can identify the ${\mathbf R}$ dependent part as the induced interaction $V_\text{ind}({\mathbf R})$ between
 the two ions. Note that Eq.~\eqref{EnergyShift} holds for both fermions and bosons.

For a BEC, the static density-density correlation function is  $\chi(q,0)=-4nm/(q^2+2/\xi^2)$ to leading order in $n^{1/3}a_{BB}$.
Using this in Eq.~\eqref{EnergyShift}, one finds~\cite{SM}
\begin{equation}
V_{\mathrm{ind}}(R)
=\frac{mV_{q=0}}{2\pi a_{BB}}\frac{\alpha}{R^4}
\label{LongrangeBECWeak}
\end{equation}
for 
$R\gg b, c, \xi$.
Thus, the long range induced interaction is  proportional to $1/R^4$ like the bare atom-ion interaction with a magnitude given by
 $\sim(m\alpha^2\pi)/(2a_{BB}r_{\mathrm{ion}}R^4)\sim (mr_\text{ion}a_{BB})^{-1}(r_\text{ion}/R)^4$ where we have used  $V_{q=0}\sim \alpha\pi^2/r_\text{ion}$.
It  is also inversely proportional to the Bose-Bose scattering length $a_{BB}$,
 reflecting that  a more compressible BEC leads to a stronger induced interaction.

For shorter distance with $b, c\ll R \lesssim \xi$, Eq.~\eqref{EnergyShift} can also be evaluated analytically giving~\cite{SM}
\begin{equation}
V_{\mathrm{ind}}(R)=
-\pi^3nm{\alpha}^2\frac{\left(b^2+c^2\right)^2}{b^2\left(b^2-c^2\right)^2}\frac{1}{R}e^{-\sqrt{2}R/\xi}.
\label{ShortrangeBECWeak}
\end{equation}
This has the same functional form as the  Yukawa interaction obtained for neutral impurities in a BEC~\cite{camacho2018landau,Camacho-Guardian2018}.

\paragraph{BEC and strong interaction.--}
For strong atom-ion interactions, we must include the atom-ion scattering matrix  in the induced interaction between the
two ions. The dominant contribution  
  is the exchange of a sound mode  in the BEC, see   Fig.~\ref{FeynFig}(b). This  gives~\cite{SM}
 \begin{equation}
V_\text{ind}(R)=\sum_{\mathbf{k}}{\mathcal T}({\mathbf k},0;0)^2\chi({\mathbf k},0)\cos({\mathbf k}\cdot {\mathbf R})
\label{VindStrong}
\end{equation}
where we have used the symmetry  ${\mathcal T}({\mathbf k},{\mathbf k}';\omega)={\mathcal T}({\mathbf k}',{\mathbf k};\omega)$.
Comparing to Eq.~\eqref{EnergyShift}, we see that the strong coupling result is obtained by substituting the bare atom-ion interaction by the scattering matrix.

When  $R\gg [b, c, \xi, m\mathcal{T}(0,0;0)]$,   Eq.~\eqref{VindStrong} gives~\cite{SM}
\begin{equation}
V_\text{ind}(R)=\frac{ m\mathcal{T}(0,0;0)}{2\pi a_{BB}}\frac{\alpha}{R^4}.
\label{VBoseStrongLongrange}
\end{equation}
 Hence, the induced  interaction is proportional to $1/R^4$ as for the weak interaction case, when $R$ is much larger than the characteristic length scale $r_\text{ion}$
 and the spatial size of the bound state that emerges at resonance. It follows from  Eq.~\eqref{VBoseStrongLongrange} that
 the interaction is very strong close to resonance where
 a new atom-ion dimer state becomes stable and ${\mathcal T}(0,0;0)$ diverges.
  In addition, Eq.~\eqref{VBoseStrongLongrange} shows that the sign of $V_{\mathrm{ind}}\left(R\right)$ is determined by $\mathcal{T}(0,0;0)$.

In Fig.~\ref{Fig-setup}, the induced  interaction potential between two ions in a BEC with  density $nr_{\mathrm{ion}}^3=1$,  healing length
 $\xi/r_\text{ion}=2$, and  different values of $b$ (or $a_{BI}$)
  is plotted. $V_{\mathrm{ind}}(R)$  increases with increasing depth of the atom-ion interaction potential (decreasing $b$), except close to the nodes, becoming very large as the resonance value
  $b\simeq0.5168r_{\mathrm{ion}}$  is approached and a bound atom-ion state
emerges.
Interestingly, $V_\text{ind}(R)$ has a node  when $b/r_\text{ion}<0.5168$ so that it is repulsive in the long range  limit, since
 $\mathcal{T}(0,0;0)$ changes sign when a bound state enters the potential, see Eq.~\eqref{VBoseStrongLongrange}.
  Figure \ref{Fig-setup} clearly shows how the strength and sign of the induced interaction depend critically on  the shape of the
 atom-ion interaction and the presence of 
a bound state.

\paragraph{Fermi gas and weak interaction.--}
We now turn to the case of two ions in a single component Fermi gas exploring first  weak atom-ion interactions so that  Eq.~\eqref{EnergyShift} is valid.
The density-density correlation function of a Fermi gas is
$\chi(q,0)=\sum_{\mathbf k}(f_{\mathbf k}-f_{{\mathbf k+\mathbf q}})/(\xi_{\mathbf k}-\xi_{\mathbf k+\mathbf q})$~\cite{Lindhard1954},
where $f_{\mathbf k}=[\exp(\beta\xi_{\mathbf{k}})+1]^{-1}$ is the Fermi function
 and $\xi_{\mathbf{k}}=k^2/2m-\epsilon_F$.
Using this in Eq.~\eqref{EnergyShift}, we can  derive~\cite{SM}
\begin{align}
V_{\mathrm{ind}}(R)=\begin{cases}
\frac{mk_F}{\pi^2}V_{q=0}\frac{\alpha}{R^4}&b\gg k_F^{-1}\\
\gamma\frac{2k_FR \cos(2k_FR)-\sin(2k_FR)}{R^4}&b, c\ll k_F^{-1}
\end{cases}
\label{FermiWeak}
\end{align}
with 
 $\gamma=\pi m \alpha^2(b^2+c^2)^2/16b^2(b^2-c^2)^{2}$ for
$R\gg b, c$. Note that the second line is of the RKKY form, which is the same as the
mediated interaction  between two impurities in a Fermi gas where the impurity-fermion interaction is short range~\cite{Ruderman1954,Kasuya1956,Yosida1957}. This
shows that the atom-ion interaction can be treated as short range when the typical interparticle spacing is
  larger than   $r_{\mathrm{ion}}$. In the high density
  regime on the other hand, the interaction has the same $1/R^4$ form as for a BEC, Eq.~\eqref{LongrangeBECWeak}, although its
  strength is reduced by $2k_Fa_{BB}/\pi\ll1$  reflecting  that the Fermi gas is much less compressible than the BEC.


\paragraph{Fermi gas and strong  interaction.--}
For strong atom-ion interaction, we 	again  include the  scattering matrix in the induced interaction.
The exchange of particle-hole  excitations in the Fermi gas between the ions give~\cite{SM}
\begin{align}
V_\text{ind}(R)=\sum_{\mathbf{q},\mathbf{k}}\left[2\theta(k_F-k)\frac{\text{Re}[\mathcal{T}^2(\mathbf{k}+\mathbf{q},\mathbf{k};\xi_{\mathbf{k}})]}{\xi_{\mathbf{k}}-\xi_{\mathbf{k}+\mathbf{q}}}\right.\nonumber\\
\left.
-\int_{-\epsilon_F}^{0}\!\frac{d\omega}\pi\frac{\text{Im}\left[\mathcal{T}^2\left(\mathbf{k}+\mathbf{q},\mathbf{k};\omega+i\eta\right)\right]}{\left(\omega-\xi_{\mathbf{k}+\mathbf{q}}\right)\left(\omega-\xi_{\mathbf{k}}\right)}
\right]\cos ({\mathbf q}\cdot{\mathbf R}),
\label{FermiStrong}
\end{align}
see Fig.~\ref{FeynFig}(c). We have assumed zero temperature and  zero population of any bound states.

\begin{figure}
\centering
\includegraphics[width=0.49\textwidth]{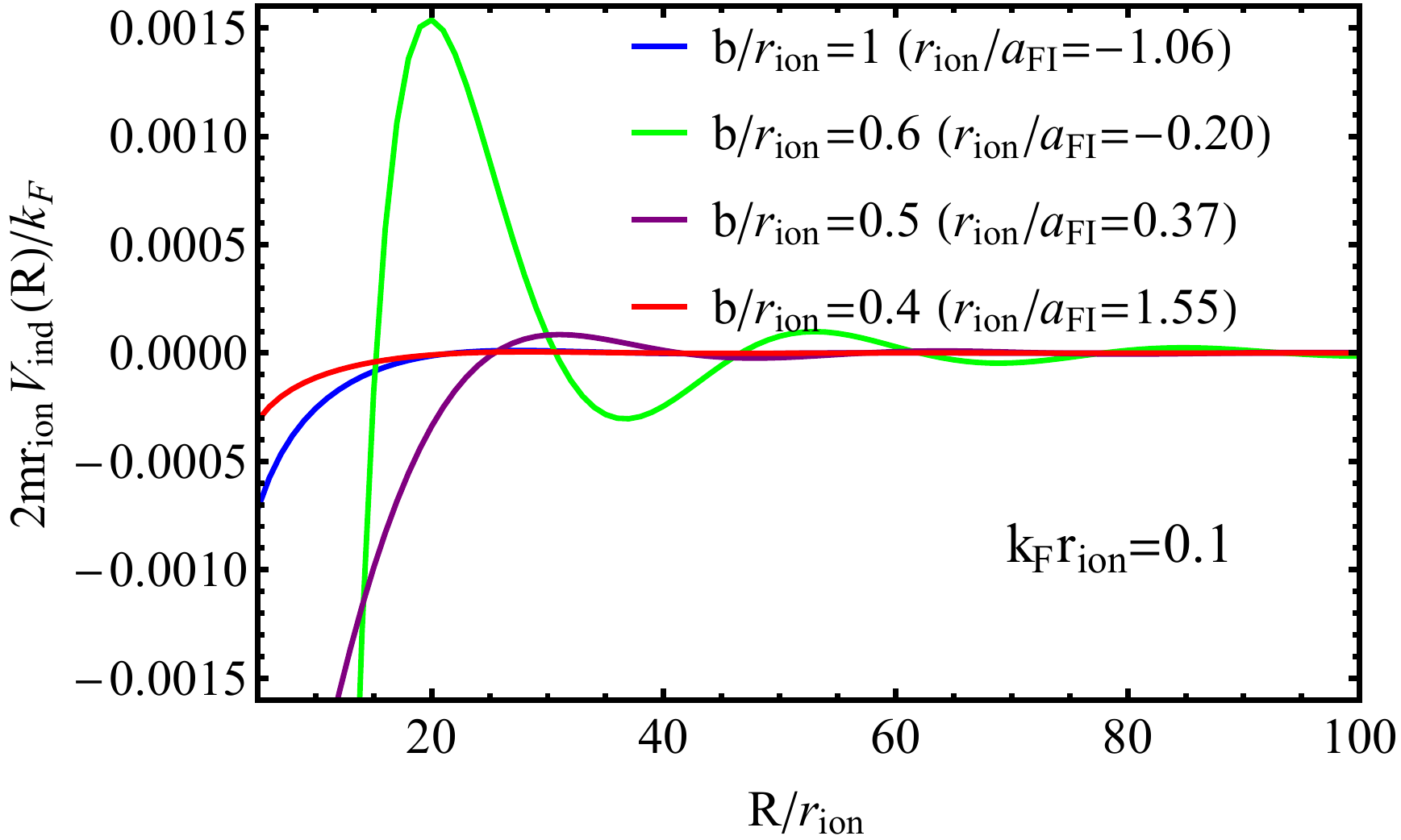}%
\caption{Induced interaction between two ions in a Fermi gas with the density $k_Fr_{\mathrm{ion}}=0.1$ and different values of  $b$ (or $a_\text{FI})$
controlling the strength of the atom-ion interaction. 
}
\label{FermiStrongFig}
\end{figure}
In Fig.~\ref{FermiStrongFig}, the induced  interaction given by Eq.~\eqref{FermiStrong} is plotted for the density $k_Fr_\text{ion}=0.1$ and
various values of $b$. Only the $s$-wave channel of the atom-ion scattering contributes for this low density,  which simplifies the numerics significantly.
Figure \ref{FermiStrongFig} shows that
the interaction increases  with decreasing $b$ towards $b=0.575r_{\mathrm{ion}}$ 
where a bound state appears at the Fermi surface. Friedel oscillations  characteristic of an interaction mediated by a Fermi sea are clearly visible. As the bound state energy
decreases  with decreasing $b<0.575r_{\mathrm{ion}}$, the interaction again decreases reflecting that it becomes off-resonant.

\paragraph{Experimental probing.--} 
We now show that the induced interaction leads to shifts  in the phonon spectrum of trapped ions.
Consider  two ions in a linear rf  trap with  
 trapping frequencies $\omega_y,\omega_z\gg \omega_x$.
  The slow dynamics in the assumed rf-field free $x$-direction is determined  by~\cite{james1998quantum}
\begin{equation}
U=\frac{\kappa}{2} (x_{2}^2+x_1^2)+\frac{Z^2e^2}{4\pi\epsilon_0}\frac{1}{x_2-x_1}
+V_\text{ind}(x_2-x_1),
\label{Vclassical}
\end{equation}
where $x_j$ is the $x$-coordinate of ion $j$ with $x_2>x_1$ and  $Ze$ is the ion charge.
 The first term  is the electrostatic trapping potential with a force constant $\kappa$, the second  is the Coulomb interaction,  and the third the
 mediated interaction between the two ions.
Even though  the equilibrium distance between the ions is  affected  by the induced interaction~\cite{SM}, it is typically
difficult to measure  without dramatically perturbing the ion-atom system.


A more promising approach is to measure the phonon spectrum of the ions with high accuracy through  a single blue sideband excitation to a
 long-lived electronic state by a narrow bandwidth laser, followed by detecting fluorescence addressing a different fast decaying transition~\cite{PoulsenPhD}.
   The frequency $\omega_x=\sqrt{\kappa/M}$ of the Kohn mode where the two ions oscillate in-phase is independent of any interaction between the ions.
  Note that the effective mass $M$ can be different from the bare ion mass due to the dressing  by the surrounding BEC~\cite{astrakharchik2021ionic,christensen2021charged}.
  This  effect is distinct from those due to the induced interaction, and it can be determined by
measuring the oscillation frequency $\omega_x$ of a \emph{single} ion -- a procedure that has  been used
for a neutral impurity in a Fermi gas~\cite{Nascimbene2009}.

\begin{figure}
\centering
\includegraphics[width=0.49\textwidth]{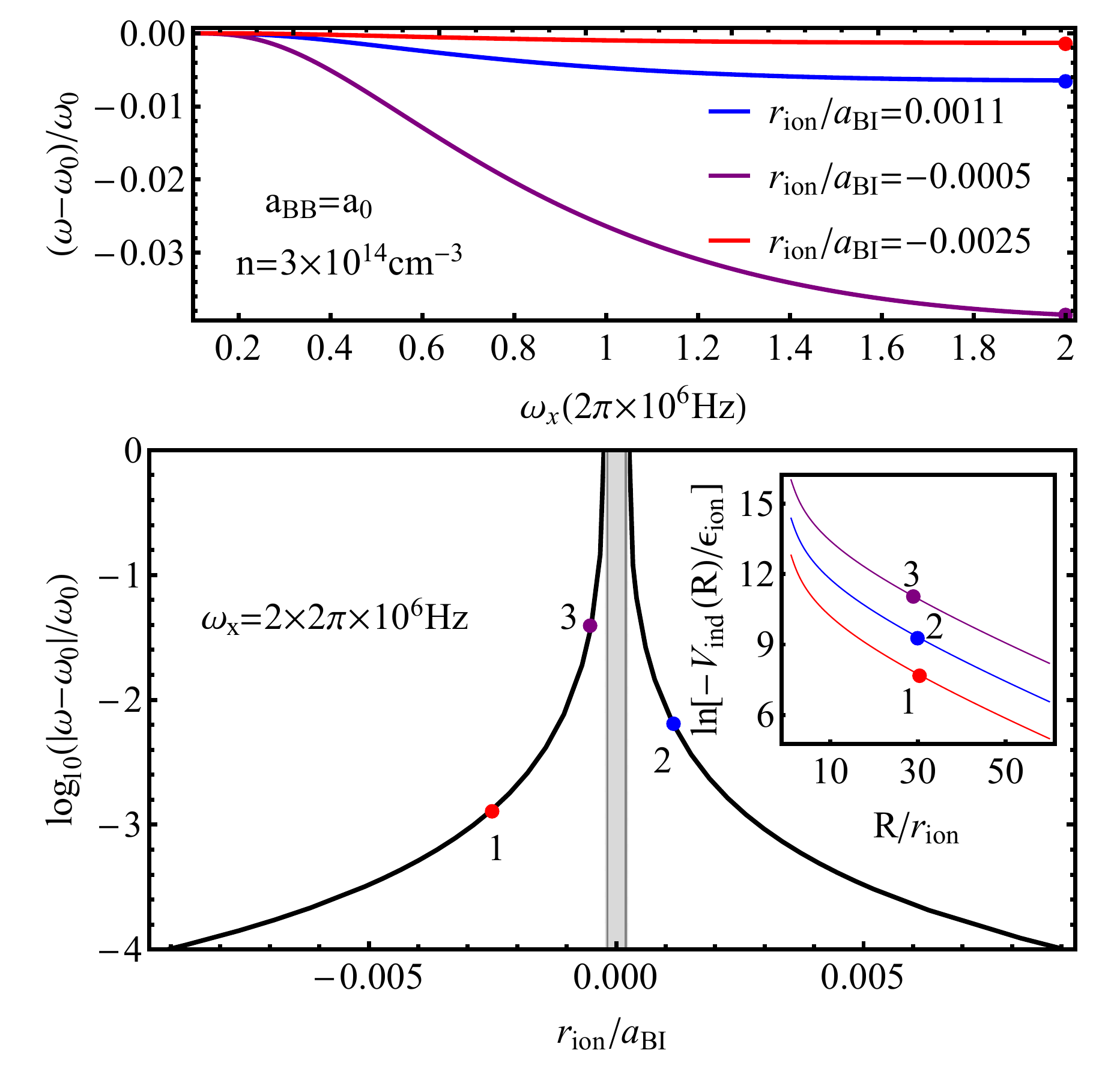}%
\caption{The relative frequency shift of the out-of-phase mode  due to the induced interaction as a function of the trap frequency $\omega_x$ (top panel),
 and as a function of the $^7$Li-$^{138}$Ba$^+$ scattering length $a_\text{BI}$ (bottom panel).
The inset shows the induced interaction for the three scattering lengths in the upper panel.
The dots show the frequency shifts and equilibrium distances (inset) for the three scattering lengths and $\omega_x=4\pi\times10^6$Hz.
}
\label{FigEquildistance}
\end{figure}
  Contrary to the Kohn mode, the mode where the two ions oscillate  out-of-phase (see the cartoon of Fig.~\ref{Fig-setup}) is affected by the induced interaction.
In the top panel of Fig.~\ref{FigEquildistance}, we plot its frequency
 obtained from Eq.~\eqref{Vclassical} by evaluating the relevant second derivatives of the induced interaction around the equilibrium positions numerically, compared to its value
  $\omega_0=\sqrt{3}\omega_x$  in the absence of the induced interaction. We use parameters
  $m$, $M=m_{\mathrm{ion}}$, and $\alpha$  appropriate for a $^7$Li-$^{138}$Ba$^+$ mixture. The BEC density  is
$n=3\times10^{14}$cm$^{-3}$, which gives $nr_{\mathrm{ion}}^3=0.136$ when $r_{\mathrm{ion}}=76.8\mathrm{nm}$. Also,
$\xi/r_{\mathrm{ion}}=20.6$ corresponding to  $a_{BB}=a_0$. This is  an approximate  value  for the $^7$Li-$^7$Li scattering length
in a wide range of magnetic fields~\cite{pollack2009extreme,Julienne2014}, where one  expects several $^7$Li-$^{138}$Ba$^+$  Feshbach resonances based on recent experimental results~\cite{weckesser2021observation}.
The  frequency shift in Fig.~\ref{FigEquildistance}  increases with trapping frequency reflecting the increased strength of the
induced interaction compared to the Coulomb repulsion for shorter  distances.
It is negative for all scattering lengths because the equilibrium distance $R_0$ between the ions is so short that the induced interaction
is attractive as seen explicitly in the inset of Fig.~\ref{FigEquildistance}. When  $R_0\gg a_{BI}>0$ and  Eq.~\eqref{VBoseStrongLongrange} holds, the induced
interaction is positive  leading to a positive frequency shift~\cite{SM}. Since we are close to resonance, this however requires a very large trap size.
A positive frequency shift for reasonable values of $R_0$ can  be obtained with a smaller $a_{BI}$  away from resonance, but it will be small and difficult
to be observed.
The shift has the largest magnitude close to $r_{\mathrm{ion}}/a_{BI}=0$ where a new atom-ion bound state appears.
 In the supplemental material, we show frequency shifts for  the case of a $^{87}$Rb-$^{87}$Rb$^+$ (or  $^{87}$Sr$^+$) mixture~\cite{SM}.

 The bottom panel of Fig. \ref{FigEquildistance} 
 shows the frequency shift as a function
of the atom-ion scattering length  $a_\text{BI}$ in the BEC using the same parameters as in the upper panel of Fig.~\ref{FigEquildistance}
and a trapping frequency  
$\omega_x=4\pi\times10^6$Hz.
Since a realistic  experimental resolution is $\Delta\omega/\omega\gtrsim10^{-4}$,  
the figure shows that the mediated interaction is observable for  $|a_{BI}|\gtrsim 100r_{\mathrm{ion}}\sim 1.5\times10^5a_0$.
The accuracy in magnetic field tuning required to achieve such a large value  depends on the width of the specific ion-atom Feshbach
resonance at hand, but assuming widths of 10-100G and typical field fluctuations  of a few mG, this should be realistic to achieve.
Indeed,  scattering lengths of the same order of magnitude have been realised  in neutral atomic gases more than a decade  ago~\cite{pollack2009extreme}.
Close to resonance where a bound state emerges and $1/a_{BI}=0$,
 the induced interaction exceeds the Coulomb repulsion leading to an instability as indicated by the
grey region in the lower panel of Fig. \ref{FigEquildistance}. 
Since the induced  interaction in a Fermi gas is weaker than in a BEC due to its smaller compressibility, it will be
 harder to observe  and may require  larger trapping frequencies.

\paragraph{Conclusions and outlook.--}
 We analysed the interaction  between two ions mediated by a BEC or a  Fermi gas. For weak atom-ion interactions, we derived several
 analytical results, and  our theory was then generalised to strong atom-ion interactions. Finally,  we
 discussed  how the control and precision of hybrid ion-atom systems can be used to  probe these mediated interactions systematically and in new regimes.

Our results motivate future work in several directions. 
It  would be interesting to explore the effects of  populating the bound states, which may give rise to
an  additional attractive interaction~\cite{Enss2020}.
Since the   micromotion energy of the ions is of the same  order of magnitude as
 the critical temperature of the BEC $T_c\sim {\mathcal O}(\mu\text{K})$, a key question concerns  temperature and heating effects.
 We expect our results to be accurate  when the condensate fraction $1-(T/T_c)^3$ is close to unity.
It is also important to explore the three-body recombination rate~\cite{Wang2017,Wang2019},
which in addition to loss also may lead to heating and molecule population. Using a Fermi gas may be advantageous for reducing such processes.
Other fascinating problems include the induced interaction in a strongly correlated Fermi gas in the BEC-BCS cross-over, and between two mobile ions.
Finally, it would be useful to
apply  approaches such as  Monte-Carlo calculations to  the  challenging strongly interacting regime~\cite{ArdilaPrivate}.

\paragraph{Acknowledgments.--} This work has been supported by the Danish National Research Foundation through the Center of Excellence (Grant agreement no.: DNRF156), the Independent Research Fund Denmark- Natural Sciences via Grant No. DFF -8021-00233B. We acknowledge useful discussions with T.~Enss, O.\ Dulieu, and X.\ Xing. 
A.\ Camacho-Guardian is thanked for providing a code for calculating the atom-ion scattering matrix.

\bibliography{dingshsh-ref}

\end{document}



\title{Supplemental Material: Mediated interactions between ions in quantum degenerate gases}
\author{Shanshan Ding}
\affiliation{Center for Complex Quantum Systems, Department of Physics and Astronomy, Aarhus University, Ny Munkegade, DK-8000 Aarhus C, Denmark. }
\author{Michael Drewsen}
\affiliation{Center for Complex Quantum Systems, Department of Physics and Astronomy, Aarhus University, Ny Munkegade, DK-8000 Aarhus C, Denmark. }
\author{Jan J.\  Arlt}
\affiliation{Center for Complex Quantum Systems, Department of Physics and Astronomy, Aarhus University, Ny Munkegade, DK-8000 Aarhus C, Denmark. }
\author{G. M. Bruun}
\email[]{bruungmb@phys.au.dk}
\affiliation{Center for Complex Quantum Systems, Department of Physics and Astronomy, Aarhus University, Ny Munkegade, DK-8000 Aarhus C, Denmark. }
\affiliation{Shenzhen Institute for Quantum Science and Engineering and Department of Physics, Southern University of Science and Technology, Shenzhen 518055, China}

\date{\today}


\maketitle


\section{Weak Interactions}
The energy shift due to two ions separated by a distance $R$ is to second order in the interaction
\begin{equation}
E_2=\sum_{M\neq G}\frac{\langle G|\sum_{\mathbf{k}',\mathbf{q}'}c^\dagger_{\mathbf{k}'}c_{\mathbf{k}'+\mathbf{q}'}V_{q'}\left(1+e^{i\mathbf{q}'\cdot\mathbf{R}}\right)|M\rangle\langle M|\sum_{\mathbf{k},\mathbf{q}}c^\dagger_{\mathbf{k}+\mathbf{q}}c_{\mathbf{k}}V_q\left(1+e^{-i\mathbf{q}\cdot\mathbf{R}}\right)|G\rangle}{E_G-E_M}
\end{equation}
where $|G\rangle$ and $|M\rangle$ are respectively the ground state and the excited state of the surrounding bath with energies  $E_G$ and $E_M$. Evaluating
this gives
\begin{equation}\begin{aligned}
E_2=\sum_{\mathbf{q}}\chi\left(\mathbf{q},0\right)V_q^2\left[1+\cos\left(\mathbf{q}\cdot\mathbf{R}\right)\right]
\label{SecondOrder}
\end{aligned}\end{equation}
where
\begin{equation}
\chi(\mathbf{q},0)=\sum_{M\neq G}\frac{2\left|\langle G|\rho_{\mathbf{q}}|M\rangle\right|^2}{E_G-E_M}
\label{BECchi}
\end{equation}
with  $\rho_{\mathbf{q}}=\sum_{\mathbf{k}}c^\dagger_{\mathbf{k}-\mathbf{q}}c_{\mathbf{k}}$ is the static density-density correlation function of the bath.
\subsection{Two ions in a BEC}
 Using that $\chi(\mathbf{q},0)=-2n_0/(\epsilon_q+2n_0g)$ for a weakly interacting BEC, the $R$-dependent part of Eq.~\eqref{SecondOrder} yields
\begin{equation}\begin{aligned}
V_{\mathrm{ind}}\left(R\right)=-\frac{2\pi^2n_0m\alpha^2}{b^2\left(b^2-c^2\right)^4}\int \left[4bc^2e^{-cq}+e^{-bq}\left(-4bc^2-b^4q+c^4q\right)\right]^2 \frac{1}{q^2+2/\xi^2} \frac{\sin qR}{qR}dq
\end{aligned}\label{Eq-fr-T-V-rfinite}\end{equation}
for the induced interaction in second order perturbation theory.

 When $R\gg b, c, \xi$, $V_{\mathrm{ind}}(R)$ can be calculated approximately with integration by parts
 $\int_a^bu(x)v'(x)dx=\left[u(x)v(x)\right]_a^b-\int_a^bu'(x)v(x)dx$ using
\begin{equation}
u_0(q)=\left[4bc^2e^{-cq}+e^{-bq}\left(-4bc^2-b^4q+c^4q\right)\right]^2 \frac{1}{q^2+2/\xi^2} \frac{1}{qR}\hspace{1cm}\text{ and } \hspace{1cm}v_0'(q)=\sin qR.
\end{equation}
This gives
\begin{equation}
V_{\mathrm{ind}}(R)=\frac{2\pi^2n_0m\alpha^2}{b^2\left(b^2-c^2\right)^4}\int dq u_0'(q)v_0(q).
\end{equation}
Another partial integration with $u_1(q)=u_0'(q)/R$ and $v_1'(q)=-\cos qR$ yields
\begin{equation}
V_{\mathrm{ind}}\left(R\right)=-\frac{2\pi^2n_0m\alpha^2}{b^2\left(b^2-c^2\right)^4}\int dq u_1'(q)v_1(q).
\end{equation}
A final partial integration with $u_2(q)=u_1'(q)/R$ and $v_2'(q)=-\sin qR$ gives
\begin{equation}
V_{\mathrm{ind}}(R)\approx -\frac{\pi m \alpha^2}{2a_{BB} b}\frac{b^2+2bc-c^2}{(b+c)^2}\frac{1}{R^4}=\frac{mV_{q=0}}{2 \pi a_{BB}}\frac{\alpha}{R^4}.
\label{Eq-fir-V-r4}
\end{equation}

When $b, c\ll R \lesssim \xi$, the exponential factor in $V_q$ can be  taken as constant $1$ in Eq.~\eqref{Eq-fr-T-V-rfinite}
because for $q\xi, qR\sim 1$, $qb, qc\ll1$. This gives
    \begin{equation}
    V_{\mathrm{ind}}\left(R\right)\approx -2\pi^2n_0m\alpha^2\frac{\left(b^4-c^4\right)^2}{b^2\left(b^2-c^2\right)^4}\int \frac{q^2}{q^2+2/\xi^2} \frac{\sin qR}{qR}dq
    =-2\pi^2n_0m\alpha^2\frac{\left(b^2+c^2\right)^2}{b^2\left(b^2-c^2\right)^2}\frac{\pi}{2R}e^{-\sqrt{2}R/\xi}.
    \label{Eq-fi-V-large-xi}\end{equation}

\begin{figure}[h]
\centering
\includegraphics[width=0.47\textwidth]{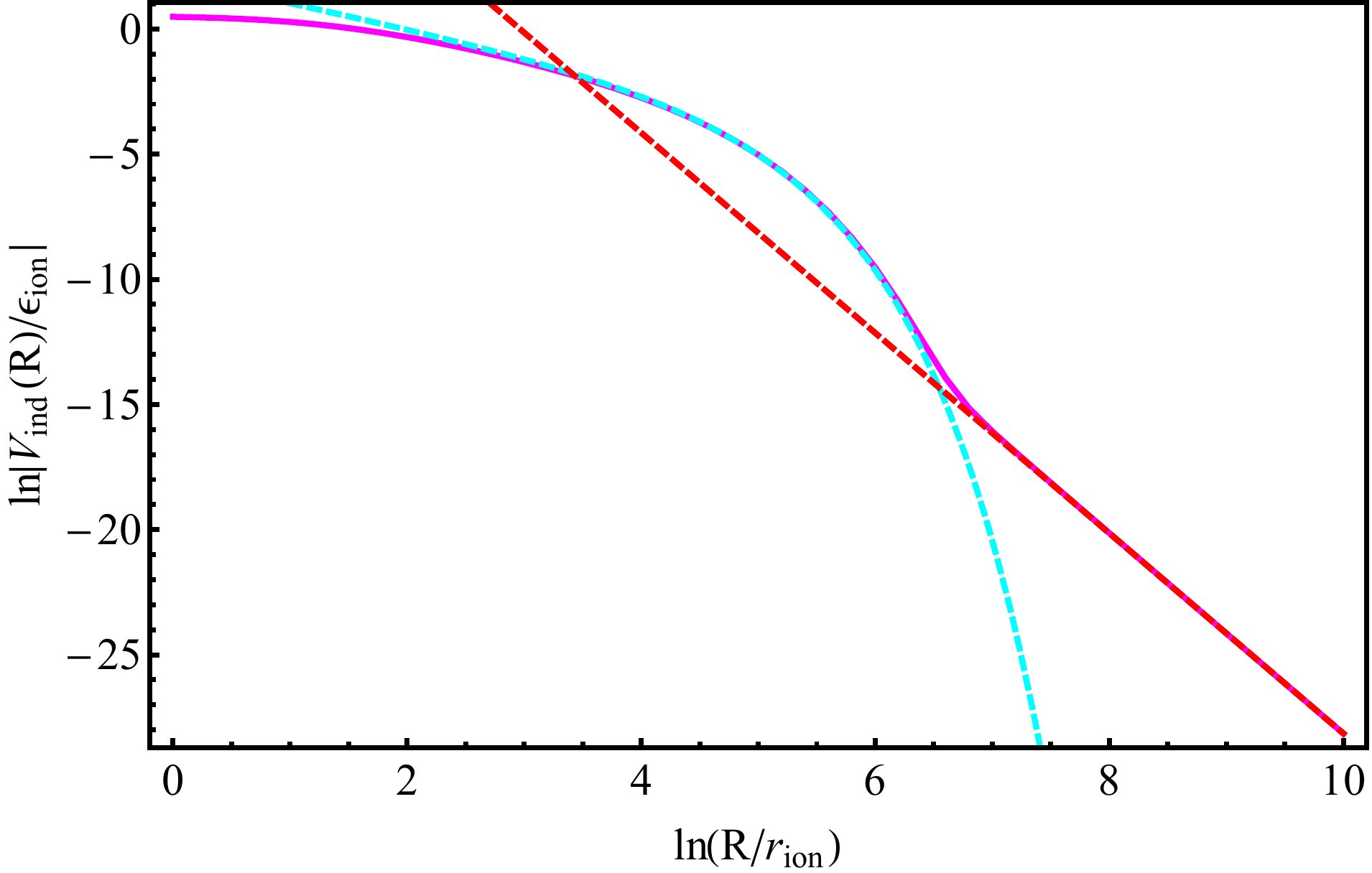}%
\caption{$V_{\mathrm{ind}}(R)$ for $b/r_{\mathrm{ion}}=1.4$ and $\xi/r_{\mathrm{ion}}=100$. The solid line is the numerical result from Eq. \ref{Eq-fr-T-V-rfinite}. The red and cyan dashed lines are respectively the analytical results given by Eqs. \ref{Eq-fir-V-r4} and \ref{Eq-fi-V-large-xi}.}
\label{Fig-V_ind-Yukawa-powerlaw}
\end{figure}

In Fig. \ref{Fig-V_ind-Yukawa-powerlaw}, we plot the induced interaction for a large healing length $\xi/r_{\mathrm{ion}}=100$ and weak atom-ion interaction with $b/r_{\mathrm{ion}}=1.4$ to illustrate that it is given by Eq. \ref{Eq-fir-V-r4} and Eq. \ref{Eq-fi-V-large-xi} for $R\gg \xi$ and $R\lesssim \xi$, respectively.

\subsection{Two ions in a Fermi gas}
For a one-component Fermi gas, the static density-density correlation function is
\begin{equation}
\chi(\mathbf{q},0)=\sum_{\mathbf{k}}\frac{f_{\mathbf{k}}-f_{\mathbf{k}+\mathbf{q}}}{\xi_{\mathbf{k}}-\xi_{\mathbf{k}+\mathbf{q}}}
=-\frac{2}{(2 \pi)^2}\frac{m}{q}\frac{1}{8}\left[4 k_F q+\left(4k_F^2-q^2\right)\ln\left(\frac{2k_F+q}{|2k_F-q|}\right)\right]
\end{equation}
where $\xi_{\mathbf{k}}=\epsilon_k-\mu$ with the chemical potential $\mu=\epsilon_{k_F}$.
Using this in Eq.\ \eqref{SecondOrder} gives
\begin{equation}
V_{\mathrm{ind}}\left(R\right)=-\frac{m}{2\left(2\pi\right)^4}\int dq V_{q}^2\frac{\sin \left(q R\right)}{R}\left[4 k_F q+\left(4k_F^2-q^2\right)\ln\left(\frac{2k_F+q}{|2k_F-q|}\right)\right].
\label{FermiSecondOrder}
\end{equation}

When $k_Fb\gg 1$, the main contribution is from $q\ll k_F$ due to the exponential decay of the potential $V_q$.
As a result, we can expand the factor in square bracket of Eq.~\eqref{FermiSecondOrder} in $\tilde{q}=q/k_F$, which yields
\begin{equation}\begin{aligned}
V_{\mathrm{ind}}\left(R\right)\approx&-\frac{m k_F^2}{2\left(2\pi\right)^4}\int dq V_{q}^2\frac{\sin \left(q R\right)}{R} 8 \tilde{q}\\
=&-\frac{m \alpha^2 k_F}{4 b^2\left(b^2-c^2\right)^4}\Bigg\{\frac{4b}{R}\Bigg[R\left(\frac{(b^4-c^4)^2}{(4b^2+R^2)^2}+\frac{2c^2(b^4-c^4)}{4b^2+R^2}+\frac{-2b^4c^2+2c^6}{(b+c)^2+R^2}\right)\\
&\qquad \qquad +4bc^4\left(\mathrm{ArcCot}\left(\frac{2b}{R}\right)+\mathrm{ArcCot}\left(\frac{2c}{R}\right)-2\mathrm{ArcTan}\left(\frac{R}{b+c}\right)\right)\Bigg]\Bigg\}.
\end{aligned}\end{equation}
The leading term for  $R\gg b, c$ is
\begin{equation}
-\frac{m \alpha^2 k_F}{b\left(b+c\right)^2}(b^2+2bc-c^2)\frac{1}{R^4}=\frac{m k_F}{\pi^2}V_{q=0}\frac{\alpha}{R^4}.
\label{Eq-fi-V-large-kF-r}
\end{equation}

 When $k_F b\ll 1$, $k_F c\ll 1$ and $R\gg b, c$, the exponential function in $V_q$ can be approximated by $1$ in Eq.~\eqref{FermiSecondOrder}
  because for $q\sim k_F$ or $qR\sim1$, $qb, qc\ll1$. We then get
\begin{equation}\begin{aligned}
V_{\mathrm{ind}}\left(R\right)\approx&-\frac{m \alpha^2 k_F^4}{2^5}\left[\frac{b^2+c^2}{b\left(b^2-c^2\right)}\right]^2\int d\tilde{q} \frac{\sin \left(\tilde{q} \tilde{R}\right)}{\tilde{R}}\left[4 \tilde{q}+\left(4-\tilde{q}^2\right)\ln\left(\frac{2+\tilde{q}}{|2-\tilde{q}|}\right)\right]\\
=&-\frac{m \alpha^2 }{2^5}\left[\frac{b^2+c^2}{b\left(b^2-c^2\right)}\right]^2\frac{2\pi}{R^4}\left[-2k_FR \cos(2k_FR)+\sin(2k_FR)\right]
\end{aligned}\label{Eq-fi-V-small-kF}\end{equation}
where $\tilde{q}=q/k_F$ and $\tilde{R}=k_FR$.

\section{Strong atom-ion interactions in a BEC}
\begin{figure}[h]
\centering
\includegraphics[width=0.5\textwidth]{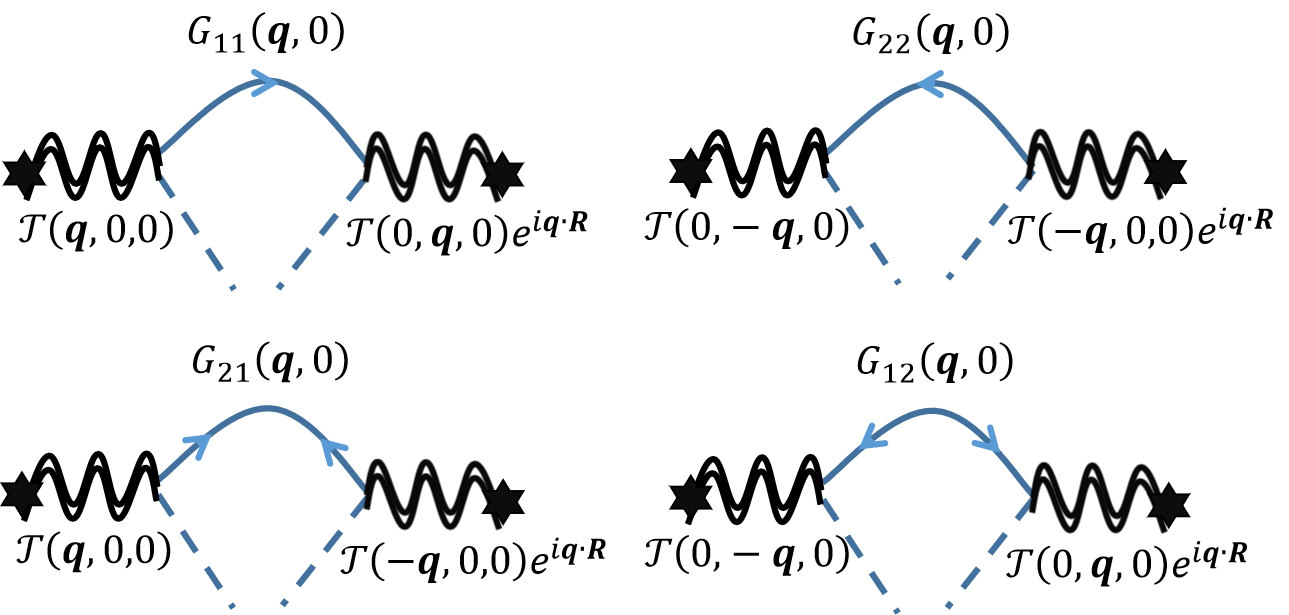}%
\caption{The diagrams for the induced interaction between two ions in a BEC.}
\label{Fig-BEC-Vind-diagram}
\end{figure}

Diagram Fig. 2b in the main text for two ions in a BEC corresponds to the summation of the four diagrams in Fig. \ref{Fig-BEC-Vind-diagram}.
This gives
\begin{equation}\begin{aligned}
V_{\mathrm{ind}}\left(R\right)=\sum_\mathbf{q}\exp\left(i \mathbf{q} \cdot \mathbf{R}\right)n_0\big[&\mathcal{T}\left(\mathbf{q},0,0\right)G_{11}\left(\mathbf{q}, 0\right)\mathcal{T}\left(0,\mathbf{q},0\right)+\mathcal{T}\left(0,-\mathbf{q},0\right)G_{22}\left(\mathbf{q},0\right)\mathcal{T}\left(-\mathbf{q},0,0\right)\\
+&\mathcal{T}\left(\mathbf{q},0,0\right)G_{21}\left(\mathbf{q},0\right)\mathcal{T}\left(-\mathbf{q},0,0\right)+\mathcal{T}\left(0,-\mathbf{q},0\right)G_{12}\left(\mathbf{q},0\right)\mathcal{T}\left(0,\mathbf{q},0\right)\big]
\end{aligned}\end{equation}
where the BEC Green's functions are
\begin{subequations}
\begin{equation}
G_{22}\left(\mathbf{k},i\omega\right)=G_{11}\left(\mathbf{k},-i\omega\right)=\frac{u_\mathbf{k}^2}{-i\omega-E_\mathbf{k}}-\frac{v_\mathbf{k}^2}{-i\omega+E_\mathbf{k}}
\hspace{0.5cm}\text{and}\hspace{0.5cm}
G_{12}\left(\mathbf{k},i\omega\right)=G_{21}\left(\mathbf{k},i\omega\right)=\frac{u_\mathbf{k}v_\mathbf{k}}{i\omega+E_\mathbf{k}}
-\frac{u_\mathbf{k}v_\mathbf{k}}{i\omega-E_\mathbf{k}}.
\end{equation}
\end{subequations}
Since $V_{\mathbf{p}'-\mathbf{p}}$ is a symmetric matrix, we have $\mathcal{T}(\mathbf{q},0,0)=\mathcal{T}(0,\mathbf{q},0)$. Using this and
$\mathcal{T}\left(-\mathbf{q},0,0\right)=\mathcal{T}\left(\mathbf{q},0,0\right)$ gives
\begin{equation}
V_{\mathrm{ind}}\left(R\right)=\sum_\mathbf{q}2\cos\left(\mathbf{q} \cdot \mathbf{R}\right)\mathcal{T}\left(\mathbf{q},0,0\right)^2n_0\left[G_{11}\left(\mathbf{q}, 0\right)+G_{21}\left(\mathbf{q},0\right)\right]=\sum_\mathbf{q}\cos\left(\mathbf{q} \cdot \mathbf{R}\right)\mathcal{T}\left(\mathbf{q},0,0\right)^2\chi\left(\mathbf{q},0\right).
\label{Eq-Vind-BEC-strong}
\end{equation}
Plugging in the static density-density correlation function of a BEC gives
\begin{equation}
V_{\mathrm{ind}}\left(R\right)=-\frac{2n_0m\alpha^2}{\pi^2r_{\mathrm{ion}}^3}\int_0^\infty \tilde{\mathcal{T}}\left(\tilde{q},0,0\right)^2 \frac{1}{\tilde{q}^2+2\left(1/\tilde{\xi}\right)^2} \frac{\sin \tilde{q}\tilde{R}}{\tilde{q}\tilde{R}}\tilde{q}^2d\tilde{q}
\label{Eq-fr-T-rfinite}\end{equation}
where $\tilde{q}=qr_{\mathrm{ion}}, \tilde{R}=R/r_{\mathrm{ion}}, \tilde{\xi}=\xi/r_{\mathrm{ion}}$ and $\tilde{\mathcal{T}}=\mathcal{T}r_{\mathrm{ion}}/\alpha$.

Given that ${\mathcal{T}}(q,0,0)$ is smooth, we can analyze the large $R$ behaviour following the same steps as for the weak interaction case above.
Using that ${\mathcal{T}}(q,0,0)$ and all its derivatives go to  $0$ for $q\rightarrow \infty$ except  on resonance, one can derive the leading term for
large $R$ as
\begin{equation}
V_{\mathrm{ind}}\left(R\right)=\frac{2n_0m\alpha^2}{\pi^2r_{\mathrm{ion}}^3}\left[\frac{2 \tilde{\xi}^2\tilde{\mathcal{T}}\left(0,0,0\right) \lim_{\tilde{q}
\rightarrow 0}\left[\tilde{\mathcal{T}}'\left(\tilde{q},0,0\right)\right]}{\tilde{R}^4}\right]=\frac{2}{\pi^2}\frac{ \mathcal{T}\left(0,0,0\right)}{g}\lim_{q\rightarrow 0}\left[\mathcal{T}'\left(q,0,0\right)\right]\frac{1}{R^4}.
\end{equation}

The derivative of the scattering matrix is from Eq.~(6) in the main manuscript
\begin{equation}\begin{aligned}
\lim_{q\rightarrow 0}\mathcal{T}'\left(q,0,0\right)=&\lim_{q\rightarrow 0}V'_q+\frac{1}{2\pi^2}\int_0^\infty k^2 d k V_k\left(-\frac{u_{\mathbf{k}}^2+v_\mathbf{k}^2}{E_{\mathbf{k}}}\right)\lim_{q\rightarrow 0}V_0'\left(q,k\right)\\
&+\sum_{\mathbf{k_1}}V_{k_1}\left(-\frac{u_{\mathbf{k}_1}^2+v_{\mathbf{k}_1}^2}{E_{\mathbf{k}_1}}\right)\frac{1}{2\pi^2}\int_0^\infty k_2^2 d k_2 V_{\mathbf{k}_2-\mathbf{k}_1}\left(-\frac{u_{\mathbf{k}_2}^2+v_{\mathbf{k}_2}^2}{E_{\mathbf{k}_2}}\right)\lim_{q\rightarrow 0}V_0'\left(q,k_2\right)+\ldots
\end{aligned}\label{Eq-dT0-dq}
\end{equation}
where $V_0(p',p)=(1/2)\int_{-1}^1 V_{\mathbf{p}'-\mathbf{p}}d\cos \theta$ ($\theta=<\hat{\mathbf{p}}',\hat{\mathbf{p}}>$) is the angular average of the interaction.
We have $\lim_{q\rightarrow 0}V'(q)=\alpha \pi^2$. Moreover, since $\lim_{q\rightarrow 0}V_0'(q,k)=\alpha \pi^2\theta(q-k)$ it follows that the second term in
Eq.~\eqref{Eq-dT0-dq} vanishes for $q\rightarrow 0$. One can show that also all higher order terms vanish.
As a result, $\lim_{q\rightarrow 0}\mathcal{T}'\left(q,0,0\right)= \alpha \pi^2$ and consequently,
\begin{equation}\begin{aligned}
V_{\mathrm{ind}}\left(R\right)=\frac{2 \mathcal{T}\left(0,0,0\right)}{g}\frac{\alpha}{R^4}.
\end{aligned}\end{equation}

\section{Strong atom-ion interactions in a Fermi gas}
Diagram Fig.\ 2c in the main text for two ions in a Fermi gas  corresponds to
\begin{align}
V_{\mathrm{ind}}\left(R\right)&=\sum_\mathbf{q}\sum_\mathbf{k}\frac{1}{\beta}\sum_{i\omega}\mathcal{T}\left(\mathbf{k}+\mathbf{q},\mathbf{k},i\omega\right)G\left(\mathbf{k}+\mathbf{q},i\omega\right)G\left(\mathbf{k},i\omega\right)\mathcal{T}\left(\mathbf{k},\mathbf{k}+\mathbf{q},i\omega\right)e^{i\mathbf{q}\cdot \mathbf{R}}\nonumber\\
&=\sum_\mathbf{q}\sum_\mathbf{k}\frac{1}{\beta}\sum_{i\omega}\mathcal{T}^2\left(\mathbf{k}+\mathbf{q},\mathbf{k},i\omega\right)G\left(\mathbf{k}+\mathbf{q},i\omega\right)G\left(\mathbf{k},i\omega\right)e^{i\mathbf{q}\cdot \mathbf{R}}
\end{align}
where $i\omega=i(2n+1)\pi/\beta$ with $n$ an integer and $\beta=1/k_BT$  is a fermionic Matsubara frequency. In the second line, we have used that
$V_{\mathbf{p}'-\mathbf{p}}$ is a symmetric matrix so that $\mathcal{T}(\mathbf{k}+\mathbf{q},\mathbf{k},i\omega)=\mathcal{T}(\mathbf{k},\mathbf{k}+\mathbf{q},i\omega)$.

Evaluating the Matsubara sum gives $V_{\mathrm{ind}}(R)=V_{\mathrm{ind}}^{12}(R)+V_{\mathrm{ind}}^3(R)$ with
\begin{equation}
V_{\mathrm{ind}}^{12}\left(R\right)=\sum_\mathbf{q}\sum_{\mathbf{k}\in k\leq k_F}\left\{2\mathrm{Re}\left[\mathcal{T}^2\left(\mathbf{k}+\mathbf{q},\mathbf{k},\xi_{\mathbf{k}}\right)\right]\frac{1}{\xi_{\mathbf{k}}-\xi_{\mathbf{k}+\mathbf{q}}}
\right\}e^{i\mathbf{q}\cdot \mathbf{R}}
\end{equation}
and
\begin{equation}
V_{\mathrm{ind}}^3\left(R\right)=\sum_\mathbf{q}\sum_\mathbf{k}\left\{-\frac{1}{\pi}\int_{-\mu}^{0}d\omega \mathrm{Im}\left[\mathcal{T}^2\left(\mathbf{k}+\mathbf{q},\mathbf{k},\omega+i\eta\right)\right]\frac{1}{\omega-\xi_{\mathbf{k}+\mathbf{q}}}\frac{1}{\omega-\xi_{\mathbf{k}}}
\right\}e^{i\mathbf{q}\cdot \mathbf{R}}
\end{equation}
in the zero temperature limit, where the integral over $\omega$ starts from $-\mu$ because $\mathrm{Im}\left[\mathcal{T}^2\left(\mathbf{k}+\mathbf{q},\mathbf{k},\omega+i\eta\right)\right]=0$ for $\omega<-\mu$.
Here, we have assumed that bound states are not populated.

\section{Two ions in a linear rf trap}
\subsection{$^{7}$Li-$^{138}$Ba$^+$ mixture}
\begin{figure}
\centering
\includegraphics[width=0.45\textwidth]{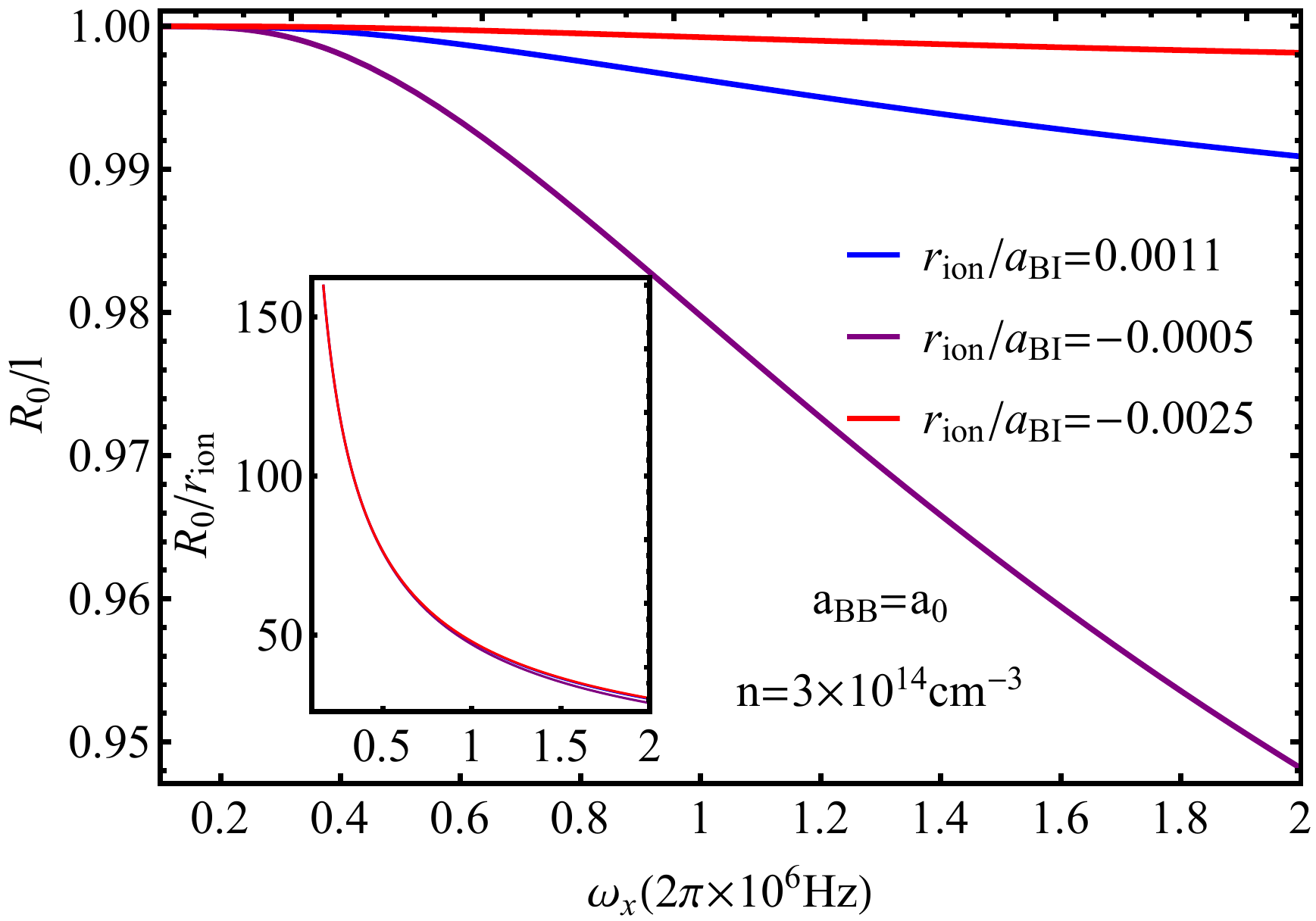}%
\caption{The ratio of the equilibrium distance between two ions with ($R_0$) and without ($l$) the induced interaction as a function of the trapping frequency $\omega_x$ for a $^{7}$Li-$^{138}$Ba$^+$ mixture. The inset shows the equilibrium distance $R_0$ as a function of  $\omega_x$.} 
\label{Fig-R0-Li-Ba}
\end{figure}

We show in Figure. \ref{Fig-R0-Li-Ba} the equilibrium distance between two ions in a linear rf trap with an induced interaction using  parameters appropriate for a $^{7}$Li-$^{138}$Ba$^+$ mixture.
\subsection{$^{87}$Rb-$^{87}$Rb$^+$ (or $^{87}$Sr$^+$) mixture}
\begin{figure}
\centering
\includegraphics[width=0.45\textwidth]{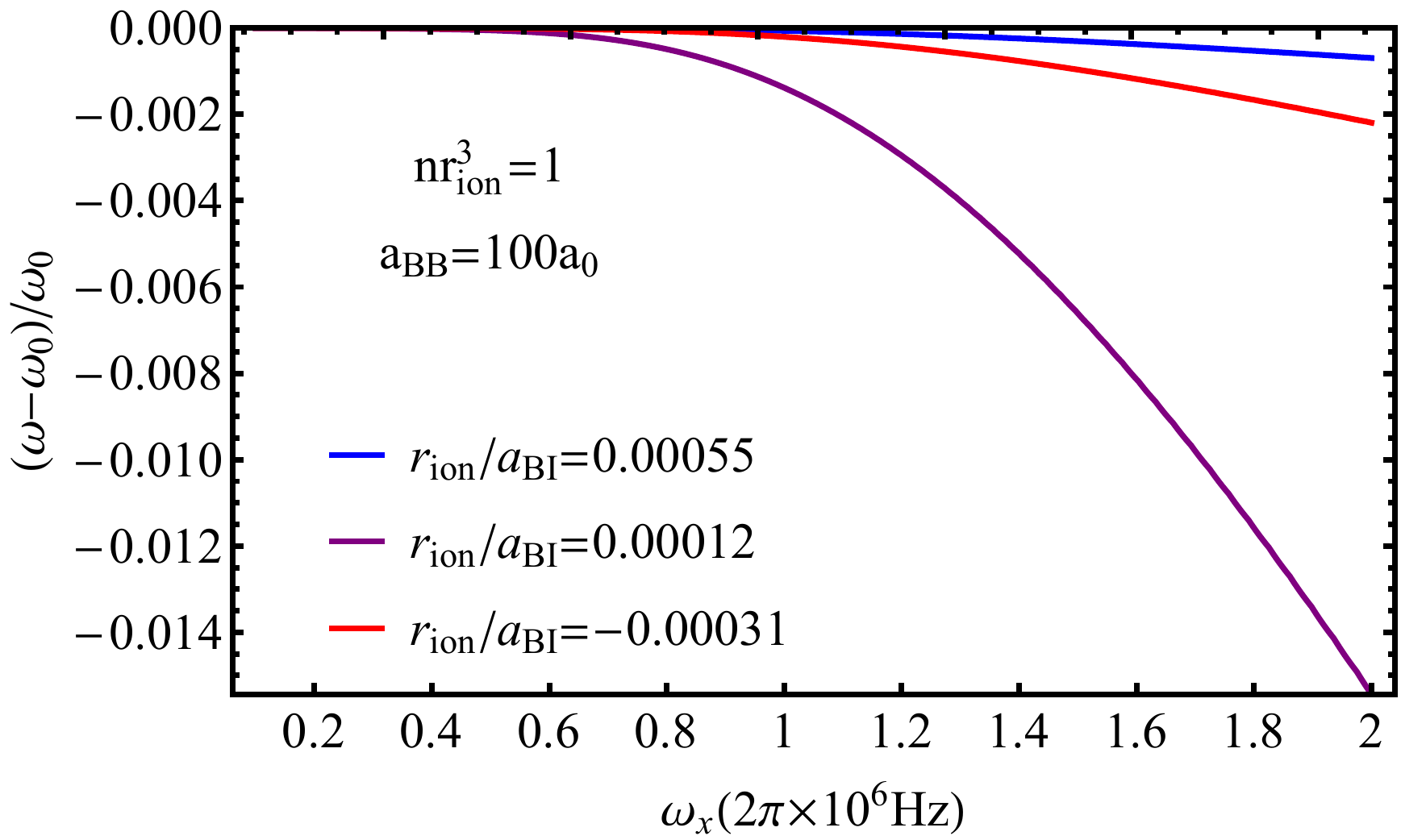}%
\caption{The relative frequency shift of the out-of-phase phonon mode for a $^{87}$Rb-$^{87}$Rb$^+$ (or $^{87}$Sr$^+$) mixture.} 
\label{Fig-frequency-shift-Rb-Rb}
\end{figure}

\begin{figure}
\centering
\includegraphics[width=0.45\textwidth]{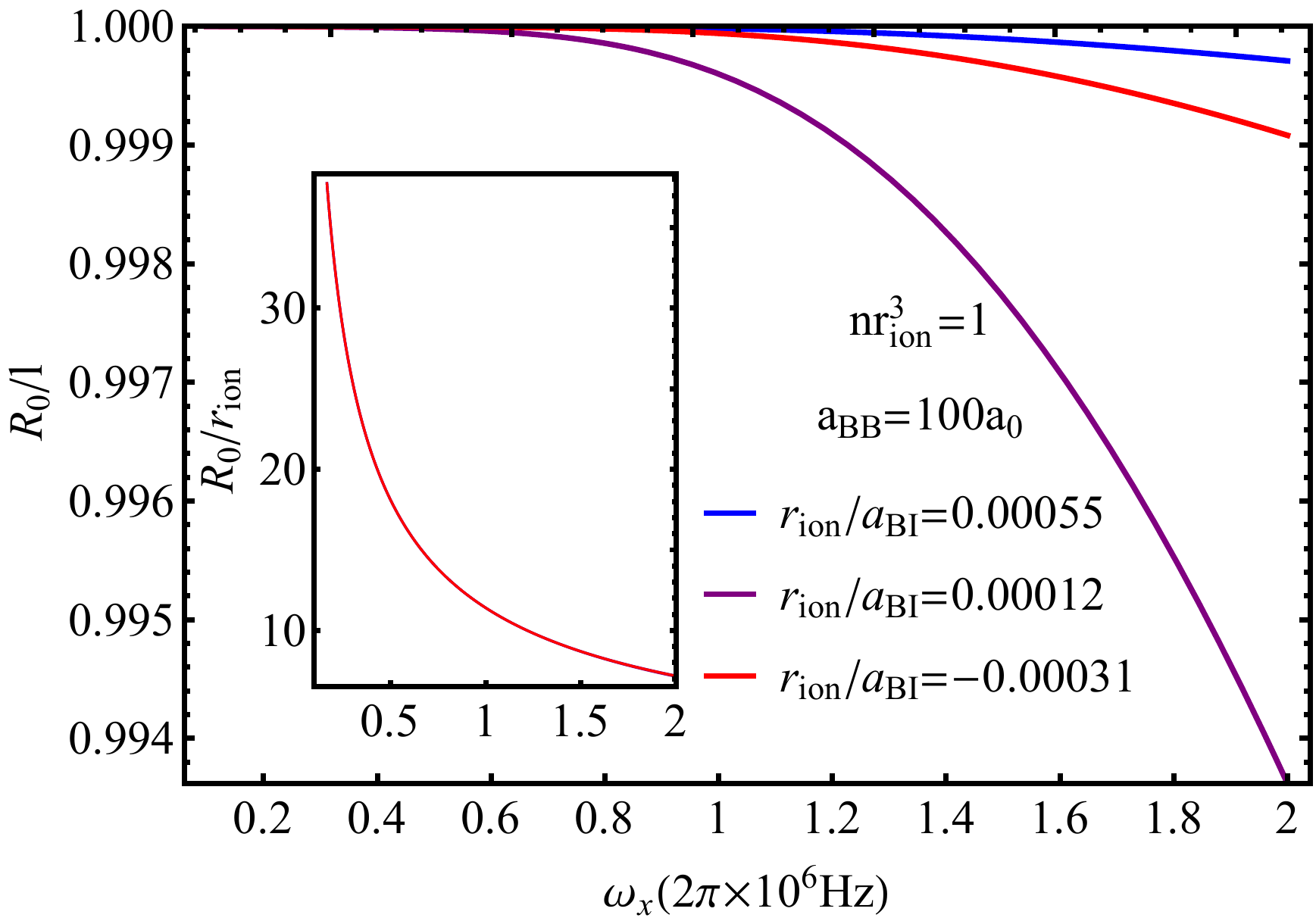}%
\caption{Same as in Fig.~\ref{Fig-R0-Li-Ba} but for  parameters corresponding to a  $^{87}$Rb-$^{87}$Rb$^+$ (or $^{87}$Sr$^+$) mixture.}
\label{Fig-R0-Rb-Rb}
\end{figure}

Figure \ref{Fig-frequency-shift-Rb-Rb} shows the  frequency shift of the out-of-phase phonon mode with the induced interaction relative to without the induced interaction using parameters appropriate for a $^{87}$Rb-$^{87}$Rb$^+$ (or $^{87}$Sr$^+$) mixture.
Figure \ref{Fig-R0-Rb-Rb} shows the corresponding equilibrium distance between the two ions.

\subsection{Analytic result for an induced interaction $C/R^4$}
When the atom-ion interaction is off resonance and the trap frequency is so small that the equilibrium distance between two ions $R_0$ fulfills the condition for Eq. (8) in the main text, one can derive $R_0$ and the relative frequency shift analytically \cite{DFVQuantum}. To be concise, we write the coefficient in front of $1/R^4$ in Eq. (8) of the main text as a $R-$independent constant $C$.

By writing the $x_j$ in Eq. (11) of main text as $x_j\left(t\right)=x_j^0+u_j$, where $x_j^0$ is the equilibrium position of ion $j$ and $u_j$ is its displacement from $x_j^0$, one has $\left[\partial U/\partial x_j\right]_{x_j=x_j^0}=0$. As a result, one obtains
\begin{equation}
x_2^0=-x_1^0=\frac{l_0\left(1\pm\sqrt{1+\tilde{C}}\right)^{1/3}}{2}\quad \text{or equivalently, } \quad R_0=l_0\left(1\pm\sqrt{1+\tilde{C}}\right)^{1/3}
\end{equation}
where $l_0=\left(Z^2e^2/4\pi\epsilon_0m_{\mathrm{ion}}\omega_x^2\right)^{1/3}$ and $\tilde{C}=8C/m_{\mathrm{ion}}\omega_x^2l_0^6\geq-1$ should be satisfied, otherwise, the system will be unstable. In addition to the original branch for $C=0$ (the $+$ one), a new branch (the $-$ one) appears for $\tilde{C}<0$ which is also unstable and will be ignored.

Since the two eigenvalues of $\left[\partial^2 U/\partial x_n\partial x_m\right]_{x_n=x_n^0,x_m=x_m^0}$ are
\begin{equation}
\begin{array}{cc}
  \lambda_1=m_{\mathrm{ion}}\omega_x^2 & \text{and} \quad \lambda_2= m_{\mathrm{ion}}\omega_x^2\left[1+\frac{4}{1+\sqrt{1+\tilde{C}}}+\frac{5\tilde{C}}{\left(1+\sqrt{1+\tilde{C}}\right)^2}\right]
\end{array}
\end{equation}
corresponding to the in-phase and out-of-phase modes, respectively, one has the frequency shift of the out-of-phase phonon mode relative to the system without induced interaction
\begin{equation}\begin{aligned}
\frac{\omega-\omega_0}{\omega_0}=\sqrt{\frac{1+\tilde{C}+\sqrt{1+\tilde{C}}}{1+\tilde{C}/2+\sqrt{1+\tilde{C}}}}-1
\end{aligned}\end{equation}
Therefore, the sign of the relative frequency shift is same as the the sign of $C$ or $a_{BI}$.
Furthermore, if $\tilde{C}\ll1$, one has $\left(\omega-\omega_0\right)/\omega_0\approx \tilde{C}/8$.
\bibliographystyle{apsrev4-1}